\documentclass[multphys,vecphys]{svmult}

\usepackage{url}
\usepackage{makeidx}     
\usepackage{graphicx}    
\usepackage{multicol}    

\makeindex             

\newcommand{\mm}{$\rm \mu m$}
\begin{document}

\title*{The Future of Far-IR / Submillimeter Astrophysics 
with
Single Dish Telescopes}
\titlerunning{Future Far-IR / Submm Single-Dish Capabilities}
\author{C. Matt Bradford\inst{1}\and Jonas Zmuidzinas\inst{1}}
\authorrunning{Bradford \& Zmuidzinas}
\institute{Division of Physics, Mathematics and Astronomy, Caltech, Pasadena, CA, 91125,
\texttt{bradford@submm.caltech.edu}, \texttt{jonas@submm.caltech.edu}}

\maketitle

\section{Introduction\label{sec:introduction}}

Although far-IR -- mm-wave astronomy has now been developing for several decades, access
to this portion of the spectrum remains difficult and limited.  Relative to their
optical-wavelength counterparts, the submillimeter observatories are both small in units
of wavelength and hot in units of photon energy,
which imposes limits on their angular resolution and sensitivity.
Nevertheless, even
the modest far-IR and submillimeter observations obtained in the last two decades have
revolutionized our understanding the physics and chemistry of the ISM, the formation of
stars, and the cosmic history of these processes.  
The demonstration by COBE 
that about half of the
radiant energy released in the universe since decoupling comes to us in 
the submm/far--IR
has further underscored the scientific necessity
of observations in this wavelength regime \cite{pug96,fds00,fix98}.
The next decade promises even greater
discoveries, as new observatories and detectors are still pushing toward fundamental
limits.  The paper briefly outlines some of the exciting capabilities and
scientific roles of new single-dish 
far--IR -- mm-wave telescopes and their instruments.

\section{Detector technologies\label{sec:detect-techn}}
Over the past decade, 
one of the most important developments for
far--IR through mm--wave astrophysics has been the push toward large--format
direct--detector arrays. 
There are two key reasons
why large arrays are important: 
(1) they provide a wider instantaneous field--of--view, 
thereby increasing the mapping speed;
(2) they also allow  better rejection of the ``sky noise'' due to
atmospheric emissivity fluctuations, which is correlated across the array,
and thereby provide better per--pixel sensitivities. Put together,
these two factors provide a dramatic improvement in imaging capability.

For observations at $\rm\lambda > 200\,\mu m$, bolometers are presently the 
most useful devices.  
Over 
the last 20 years, the
mapping speed of bolometer arrays -- equal to the ratio of the number of pixels to the
square of a single pixel's sensitivity -- has been improving at about a factor of two per
year 
(faster than Moore's law !). 
Gains have been achieved in both single pixel sensitivity 
and array size.  
The individual detector elements are now
achieving background-limited performance for ground--based astronomy or CMB studies, 
and the number of 
pixels 
in a typical array is $\le 400$, 
which is about 
the practical limit of what
can be achieved for individually--wired detectors in a sub-Kelvin focal plane.  

Multiplexing technologies for the next generation of bolometer arrays are now under
development by several groups for both CMB studies and submillimeter astronomy.  The most
ambitious example at present is the SCUBA-2 effort, 
slated to deploy 5000 close-packed superconducting transition-edge detectors in each of two
(450\ $\mu$m and 850\ $\mu$m) arrays, to be read out with a SQUID-based multiplexer
\cite{hol03,dun03}.
However, while these technologies are very promising, they are in the early phases, and there remains the substantial task
of fielding robust, sensitive astronomical instruments at the telescope.

The transition from the mm/submm 
to the far-IR ($\rm\lambda < 200 \mu m$) 
is a shift from ground-based observations to airborne and orbital platforms. 
The detectors of choice thus far have been the Ge:Ga photoconductors, 
which have flight heritage, require only modest cryogenic 
infrastructure (T$\sim$1.5~K), and can be naturally employed in 
large-format arrays using cryogenic CMOS multiplexers.
These detectors arrays are the heart of the MIPS instrument on SIRTF (32$\times$32 at 70\mm, 2$\times$20 at 160\mm) 
and the PACS instrument on Herschel (16$\times$25 at 90\mm\ \& 160\mm), and will offer the most sensitive
observations in the far-IR to date. 
Nevertheless, substantial room for improvement exists even at these
shorter wavelengths:
array construction is still very labor--intensive, which limits 
the array size, 
especially for stressed detectors at $\lambda > 100\,\mu$m;
the quantum efficiencies are still relatively low;
and the temporal response of these detectors
requires sophisticated calibration protocols.

Coherent detection technology has also achieved impressive gains,
driven by the needs of projects such as HIFI/Herschel and ALMA.
Sensitive superconducting tunnel junction
(SIS) mixers now operate up to 1.25~THz (240~$\mu$m), while
superconducting hot--electron bolometer (HEB) mixers
push to higher frequencies still, beyond 2.5~THz (120~$\mu$m).
The improvements in local oscillator technology have been even more
dramatic, especially the solid--state amplifier/multiplier chains 
developed for HIFI/Herschel, which are now capable of continuous
tuning over $> 100\ $GHz bandwidths and provide ample power
for pumping SIS mixers at 1.25~THz and HEB mixers at 1.6~THz.
These developments will revolutionize high--resolution
terahertz spectroscopy, especially using airborne (SOFIA) and space (Herschel) platforms. 

\section{New single-dish facilities} 
A major initiative for the submillimeter / millimeter astrophysics community
in the next decade is the Atacama Large Millimeter Array (ALMA).  
ALMA will be an inteferometric array of 64 12-meter antennas 
operating in the 230, 345, 460, 650, and 850 GHz atmospheric
windows, and will be a very powerful instrument in these frequency bands
with its unique combination of high sensitivity and very high
spatial and spectral resolution. 
While future observatories must carefully consider their role in this context, 
the development of large--format cameras -- one of the major technological
trends over the past decade -- offers powerful
new capabilities for single--dish observatories. 
While not widely appreciated,
it is in fact possible for future single-dish facilities equipped with
wide--field direct--detector cameras to be competitive with ALMA in terms of absolute 
point--source continuum sensitivity,  and 
orders of magnitude faster than ALMA for wide--field imaging, 
thereby securing a scientifically important role which is complementary to ALMA.
\begin{figure}[t] \centering
\includegraphics*[height=12cm]{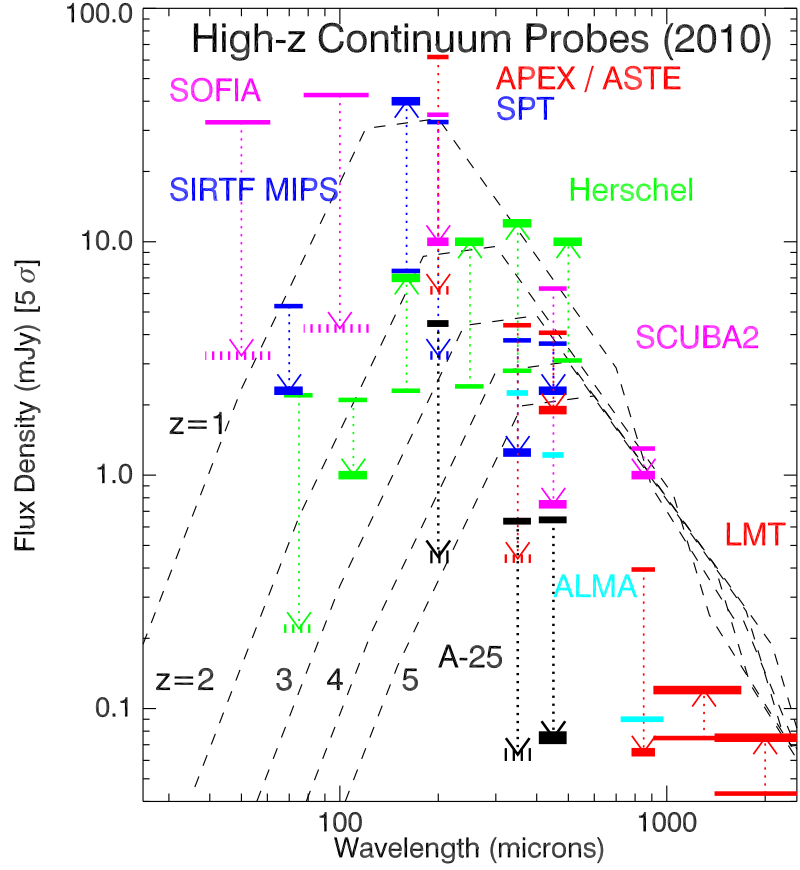}
\caption{Continuum capabilities in the far-IR .  Raw sensitivities (5$\sigma$, 1
hour) are indicated with thin horizontal lines, taken from the
references in Table~\ref{tab:1}, or calculated in the photon background limit assuming
50\% instrument transmission.  Estimates of the confusion
limit due to extragalactic sources are indicated with thick horizontal lines at the
termination of the vertical arrows.  The confusion limits are taken to be 5 $\times$ the
flux density at which there is 1 source per beam according to the source models of
Blain et al.~\cite{bla02}.  For instruments which are not confusion-limited in 100 hours, the
5$\sigma$, 100 hour sensitivity is plotted.  Upward arrows imply measurements which are
confusion-limited in less than 1 hour, downward arrows the converse.  As a guide to the
detectability of dusty galaxies, the Arp~220 SED at
redshifts from 1 to 5 is overplotted in the light curves (taking $\rm \Omega_{matter}=.38$, $\rm
\Omega_{\Lambda}=0.62$).}  
\label{fig:contsens}       
\end{figure}

Consider the achievable (background-limited) sensitivity of a observatory:
\begin{equation}
\rm NEFD_{BG} = \frac{h\nu\,
[\bar{n}(\bar{n}+1)]^{1/2}}{\eta_{inst}\,\eta_{tel}\,\eta_{atm}A_{tel}(N_{pol}\Delta\nu)^{1/2}},
\label{eq:1}\end{equation} where 
$\rm \bar{n} = \epsilon_{load}\,\eta_{inst}\,(e^{h\nu/kT}-1)^{-1}$ 
is the photon mode occupation number
(equivalent to the number of photons per second per Hz of 
detection bandwidth in a single spatial mode arriving at a detector).
Although ALMA will offer a tremendous collecting area,
single--dish telescopes can offer improvements in all the other factors appearing in
Eq.~(\ref{eq:1}).  A reduction of 
the atmospheric opacity improves the
sensitivity twice -- in increasing the transmission ($\eta_{\rm atm}$) 
and in reducing the background load ($\epsilon_{\rm load}$).  
The telescope efficiency $\eta_{\rm tel}$ is
determined by the antenna surface accuracy according to the
Ruze formula: $\rm \eta_{tel}~\approx~\eta_{0}\exp{[-(4\pi\epsilon_{rms}/\lambda)^2]}$.
The ALMA antennas will have $\rm \epsilon_{rms} \sim25~\mu m$, leaving
room for improvement at the high frequencies.  

\begin{table}[t]
\centering
\caption{Future ground-based single-dish observatories }
\label{tab:1}       
\begin{tabular}{lcccc}
\hline\noalign{\smallskip}
Observatory &\ \  Dia.\ \  &\ \  Surf RMS\ \  &\ \  PWV (top 25\%)\ \  & Refs \\
\noalign{\smallskip}\hline\noalign{\smallskip}
SCUBA2-JCMT & 15 m & 20$\rm\mu m$ & 0.7 mm & \cite{hol03} \\
APEX / ASTE & 12 m & 18$\rm\mu m$ & 0.5 mm & \cite{apex,aste} \\
SPT & 10 m & 15$\rm\mu m$ & 0.4 mm & \cite{spt} \\
LMT & 50 m & 70$\rm\mu m$ & 0.8 mm & \cite{lmt}\\
A-25 & 25 m & 12$\rm\mu m$ & 0.4 mm & (adopted)\\
\noalign{\smallskip}\hline
\end{tabular}
\end{table}
\subsection{Direct-detection continuum observations}
In addition to the sensitivity improvements that are possible by improving the telescope
accuracy and atmospheric conditions, direct-detectors can provide additional sensitivity
gains since one can couple both polarizations and the full bandwidth provided by the
atmospheric windows.  The advantage can be substantial -- the 650 and 850 GHz windows are
each 100 GHz wide, while ALMA will provide only 4--8~GHz instantaneous band.  When combined
with the efficiency factors, the per pixel sensitivity for a large single-dish telescope
can be comparable to that of ALMA.  The thin horizontal lines at the start of the vertical
arrows in Figure~\ref{fig:contsens} plot the raw sensitivities of planned and potential
observatories, listed with their parameters in Table~\ref{tab:1}.  In the short
submillimeter windows, the 10--15 meter class telescopes at excellent sites (APEX, ASTE,
SPT) offer continuum sensitivities within a factor of 2--5 of ALMA.  The fiducial A25
telescope, a 25--meter with $\rm \epsilon_{rms}=12\mu m$ at a very dry site would
be 2--5 times {\it more sensitive} than ALMA in these windows.
\begin{table}[t] \centering
\caption{Estimated far-IR / submm galaxy detection rates}
\label{tab:2}       
\begin{tabular}{lcccc}
\hline\noalign{\smallskip}
 &\ \   200\mm\ \  &\ \   450\mm\ \  &\ \   850\mm\  &\ \   1.3mm\ \    \\
\noalign{\smallskip}\hline\noalign{\smallskip}
\hspace{0.1in}$f_{nu}$, most likely [mJy]& 10--1000 & 10--50 & 2--7 & 1--5\\
\noalign{\smallskip} \noalign{\smallskip}
{\it Platform} & \multicolumn{4}{c}{{\it Galaxy Detection Rate} [hr$^{-1}$]}\\
\noalign{\smallskip}
SIRTF MIPS & 70 \\
Herschel PACS& 100 \\
SPT w/ $10^3$ beams & & 150 & & \\
APEX / ASTE (300 beams) & & 25 & & \\
SCUBA2 & & 20 & 100 & \\
LMT &  &  & 100 & 700 \\
A-25 w/ $10^3$ beams& 7 & 900 & 2000 & \\
ALMA & & 1 & 60 & 40 \\
\noalign{\smallskip}\hline
\end{tabular}
\end{table}
\subsubsection{Large format direct-detector arrays allow rapid surveys}
While the point source sensitivity of a large single-dish telescope can approach that of
ALMA at the highest frequencies, the real advantage for single-dish telescopes is the
tremendous mapping speed afforded by large-format focal plane arrays.  As pointed out
above, the mapping speed scales as the instantaneous solid angle and the inverse square of
the sensitivity.  To get a useful measure of survey speed, we convert the mapping speed to
a galaxy detection rate, using power-law extrapolations to the observed source counts in a
manner similar to \cite{bla02}.  As the estimates presented in Table~\ref{tab:2} show, a
single-dish telescope with a large detector array is much more efficient than ALMA for
discovering the sources that make up the IR background.  For example, at 450~\mm, the most
likely galaxy to be detected has a flux of a few~$\times$10~mJy.  ALMA will be capable of
detecting about one of these per hour, while SCUBA-2 will detect some 20 per hour, and a
SCUBA-2 type array on the A-25 telescope could detect $\sim10^3$ galaxies per hour.
Of course, this rapid mapping speed provided large-format detector arrays applies to
a wide range of science topics beyond the survey experiment outlined here.

\subsubsection{Source confusion is a limitation for the 3-meter class telescopes}
A key limitation of continuum surveys using single-dish telescopes is source confusion.
When the number of sources in a beam approaches a substantial fraction of unity,
extraction of fainter flux densities becomes difficult. Figure~\ref{fig:contsens} includes
estimates of this limit for the single-dish telescopes with the thick horizontal lines.
We highlight some key points: 1) SIRTF and Herschel will be quickly confusion-limited at
their longest wavelengths, and the resulting effective sensitivities are insufficient to
probe even bright ($\rm L=10^{12} L_{\odot}$) sources beyond $\rm z\sim 1$.  2)
Ground-based telescopes can in principle survey bright sources to $\rm z\sim 5$ in the 350
and 450~\mm\ windows, but with the existing and planned 10--15-meter telescopes this will
require many tens of hours of observations (of a single field) in good weather, with
careful attention paid to systematics.  The fiducial A-25 telescope would be a factor of 5
more sensitive than the other telescopes at these wavelengths, and yet effectively not
confusion-limited because of the smaller beam.  3) At 850~\mm\ and 1.3~mm, the observed flux
density is essentially independent of redshift, and beam confusion limits a 15-meter telescope
to only the bright sources, as has been found with SCUBA.  In principle, the 50-meter LMT could detect
sources more than 10$\times$ less luminous than its 15-meter counterpart at 850~\mm, but
for $\rm \lambda > 1mm$, the confusion again prohibits the detection
of fainter sources.

\subsection{Far-IR - mm spectroscopy with single-dish telescopes}
ALMA will be an exquisite platform for spectroscopy and once commissioned will be the instrument of
choice for spectroscopy at known frequencies and positions in its operational bands.
Nevertheless, there will remain important scientific niches for spectroscopy that cannot be addressed with ALMA.
\subsubsection{Full far-IR coverage provided by Herschel and SOIFA}
Perhaps the most important is the full far-IR spectral coverage provided by Herschel and
SOFIA, which will facilitate ISM studies not possible at other wavelengths.  With few
exceptions, the neutral atomic and ionized gas phases are probed with spectral lines at
$\rm\lambda < 200 \mu m$, and the optical and near--IR transitions which might be
important are typically obscured by dust.  Key tracers of the molecular phase are also
unique to the far-IR: examples include rotational transitions of the light molecular
hydrides (e.g. OH, CH, HD), mid-- and high--J CO lines which cool excited ($\rm T>100 K$)
molecular gas, and H$_2$O--- a key constituent of the molecular ISM not
observable from the ground.  It is impossible to constrain the dense ISM conditions
with ground-based observations alone, and much of our current understanding of these
regions is based on far-IR spectroscopy from ariborne and space platforms:
COBE, the KAO, ISO and SWAS.  As the sensitivity of these observations improves by two or
more orders of magnitude in the next decade with SIRTF, SOFIA and Herschel, we can
anticipate significant advances in our understanding of the dense ISM phases advances.

\begin{figure}[t] \centering
\includegraphics[height=12cm]{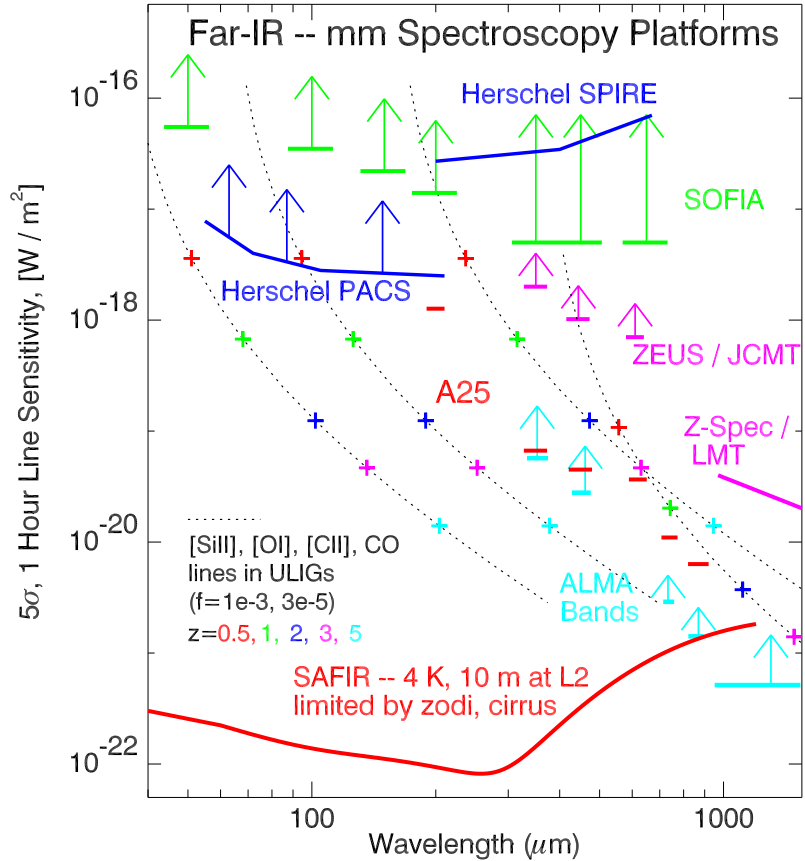}
\caption{Sensitivity provided by spectroscopy platforms in the far-IR and submillimeter.
The upward arrows account for the effective sensitivity for line surveys, incorporating
the time to cover the full atmospheric window for the ground-based instruments, or a 20\%
band for the SOFIA and Herschel instruments.  Thin dashed curves show the intensities of
diagnostic fine-structure and CO lines from high-redshift galaxies.  Sensitivities for the
Herschel and SOFIA instruments and ALMA are taken from the appropriate web pages, and Zeus
and Z-Spec are referenced in the text.  The sensitivities of SAFIR and the fiducial A25
telescope are calculated according to the photon noise limit assuming a 25\% instrument
transmission in a single polarization. 
SAFIR's background noise is due to zodiacal light and galactic
cirrus, taken in regions of low intensity.  {\it (Contact author for color version.)}}
\label{fig:specsens}       
\end{figure}
\subsubsection{Broad-bandwidth surveys with direct-detection spectrometers}
Another niche in the context of ALMA's limited bandwidth is the capability to couple a
large instantaneous band from point sources for spectral surveys.  This is both a means of
rapidly measuring several lines simultaneously, and potentially as
a redshift measurement technique for optically-faint sources.  The spectral survey rate
increases as the instantaneous bandwidth of instrument, and even from the ground the
instantaneous bandwidth can be an order of magnitude larger than the 4--8~GHz coupled by
ALMA.  It can be shown that the most sensitive approach to broad-band spectroscopy of
point sources is a low-order grating spectrometer \cite{bra03}, and for the first time such devices are
under development for the submillimeter and millimeter.  We highlight ZEUS,
an echelle spectrometer for the 350~\mm\ and 450~\mm\ windows at the JCMT \cite{nik03} and
Z-Spec, a new waveguide grating spectrometer which covers the full 200--300 GHz window
instantaneously \cite{bra03}.  In their windows, these ground-based spectrometers
are more sensitive than those of Herschel and SOFIA because of the aperture size, and should detect diagnostic lines in bright sources ($\rm L\sim few \times 10^{12}\,L_\odot$) up to z$\sim$3.
Furthermore, such instruments combined with
the A25 telescope could improve the sensitivity by an order of
magnitude -- probing the more more typical ($\rm L\sim few \times 10^{11}\,L_\odot$)
galaxies.   
\subsubsection{Potential of a cold space telescope}
With Herschel, SOFIA, and the currently--envisioned ground-based spectrometers, we will
have the capability to
study the ISM conditions in local galaxies, and the very brightest galaxies out to
redshifts of $\sim3$, but not to the earliest epochs of galaxy formation.  These platforms are
limited by the background noise from the warm telescope itself (and atmosphere).  However, dramatic
improvements in sensitivity are still possible with a space telescope cooled to $\rm
T<10\,K$.  As an example, Fig.~\ref{fig:specsens} includes the background-limited
sensitivity for SAFIR, a 4~K, 10 meter observatory recommended in the McKee/Taylor Decadal
Survey report.  SAFIR will be an excellent far-IR
spectroscopy platform -- capable of detecting spectral lines in a Milky Way--type galaxy to
redshifts of 5 or more.  Note that a spectrometer can readily detect lines from several
sources in the same beam, and thereby offers the potential to overcome the spatial
confusion limit.


%
%

%
%

\printindex
\end{document}